\RequirePackage{fix-cm}
\documentclass[onecolumn]{svjour3questa}       
\usepackage{amsmath,amssymb}
\usepackage{mathptmx}      
\usepackage{hyperref}
\usepackage[T1]{fontenc}

\def\P{\mathrm{P}}

\newcommand*\diff{\mathop{}\!\mathrm{d}}
\newcommand{\one}{\boldsymbol{1}}

\DeclareMathOperator*{\argmax}{arg\,max}

\begin{document}

\title{Queue input estimation from discrete workload observations}

\author{Liron Ravner}

\institute{Liron Ravner at
           Department of Statistics, University of Haifa, Israel.
           \email{lravner@stat.haifa.ac.il}
       }

\date{\today}

\maketitle

\section{Introduction}

This note considers an M/G/1 queue for which the workload process is observed periodically. The goal is to estimate the arrival rate $\lambda$ and the parameters of the job-size distribution $G$.  The main challenge is making optimal use of the available information, in a statistical sense.     In particular,  minimizing the variance of the estimation errors is an important open problem.  The discussion presented here is applicable to the more general framework of a L\'evy-driven queue (see \cite{DM2015}), but for the sake of brevity we focus on the special case of the M/G/1 model.  

In many queueing systems there is uncertainty regarding the distribution of the input process.  This calls for the development of statistical inference techniques for estimating the desired properties. However,  it is often hard to apply standard statistical methodology due to the intricate dependence between observations induced by the queue dynamics.  Furthermore, the primitives themselves are often not observed and indirect estimation methods are therefore required.  In our setting, periodic workload observations in an M/G/1 queue are clearly dependent and their (joint) time dependent distribution is intractable. Therefore,  estimators based on the stationary properties of the system are often used (e.g., \cite{HP2006}). Another possibility is applying conditional likelihood approaches for associated random variables with a known distribution, for example when inter-sampling times are exponential (see \cite{RBM2019}).  See Asanjarani et al.  \cite{ANT2021} for an in depth review of the literature on statistical inference for a wide range of queueing systems and observation schemes, and Sections~2.4,3.4 in particular for the dynamics of periodic queue/workload observations.  Our aim is estimation of the input parameters for any discrete sampling scheme, and ideally to do so in a way that is as efficient as possible, i.e., derive a consistent estimator that minimizes the asymptotic variance of the estimation errors.  


\section{Problem statement}

Jobs arrive to a FCFS queue according to a Poisson process $(N_t)_{t\geq 0}$ with rate $\lambda>0$. The job sizes $(B_j)_{j\geq 1}$ are an iid sequence with a common distribution $G(x)=\P(B\leq x)$. Work is processed at a unit rate and the workload at time $t$ is
$
W_t=W_0+\sum_{j=1}^{N_t}B_j-\int_0^t\one\{W_u>0\}\diff u$.
Denote the net input to the system by $X_t=\sum_{j=1}^{N_t}B_j-t$, then $(W_t)_{t\geq 0}$ is the net input process reflected at zero.  The distribution of the compound Poisson input  is determined by a finite collection of parameters $\theta\in\Theta\subseteq\mathbb{R}^p$. E.g.,  in an M/M/1 queue the parameters are the arrival and service rates: $\theta=(\lambda,\mu)$. 

The workload is sampled at a collection of times $\mathrm{T}_n=(T_1,\ldots,T_n)$.  The goal is to estimate $\theta$ from the sample $\mathrm{W}=(W_{T_1},\ldots,W_{T_n})$. A natural approach is to compute the maximum likelihood estimator (MLE): 
$\hat{\theta}_n=\argmax_{\theta\in\Theta}L(\theta;\mathrm{W})$,
where $L(\theta;\mathrm{W}):=\diff\P_\theta(W_1,\ldots,W_n)$ is the joint distribution of the workload samples (which is continuous on the real line and has an atom at zero). Under some regularity conditions the MLE is known to be consistent and statistically efficient; as $n\to\infty$, the MLE converges in probability to the true parameter and minimizes the asymptotic variance of the estimation errors. The problem is that the joint distribution $\P_\theta(w_1,\ldots,w_n)$ of workload observations is intractable.   This calls for development of alternative estimation techniques that either approximate the likelihood or utilize the available data in a different manner.  The question is then, what is the best way, in a statistical sense, to make use of the periodic workload observations?

\section{Discussion}

For the stationary workload distribution, closed form expressions of the moments and the Laplace-Stieltjes transform (LST) are available. Intuitively, if the sampling times are far enough apart the sample can be treated as iid sequence of stationary workload observations: $\P_\theta(w_1,\ldots,w_n)\approx\prod_{i=1}^n\tilde{\P}_\theta(w_i)$, where $\tilde{\P}$ is the stationary workload distribution. Note that the empirical LST needs to be inverted in order to compute the likelihood function (see  \cite{BKV2009}),  which adds additional noise to the estimators. Other stationary properties of an M/G/1 queue are also useful for estimation of the input parameters.  In \cite{HP2006} the relation between the stationary workload and the residual job-size distribution is utilized for nonparametric inference of $G$.  This approach yields a consistent estimator with asymptotically Gaussian errors, however it entails several technical difficulties associated with the residual job-size distribution and does not perform well in the extreme cases of lightly or heavily loaded systems. 
Diffusion approximations of highly loaded systems are also used for estimating the input parameters (see \cite{RTP2007}).  The diffusion approximation further has the desirable feature that the correlation structure of the transient observations is known. 
 
The sampling times need not be deterministic, and it turns out that sampling the queue at random times according to an external Poisson process yields many estimation methods.  This is known as Poisson probing (e.g.,  \cite{NKS2009,NW2009,AJP2014}). While estimators devised using Poisson probing of the workload often rely on stationary properties (such as moments and transforms), the correlation structure of the observations and its implications on the estimation accuracy is taken into account. Poisson probing further enables capturing the dependence between consecutive observations directly. In \cite{CWM1994} the likelihood is replaced by a recursive approximation of the workload density conditional on the previous workload observation.  Another approach relies on the known transient (LST) of the workload at exponential sampling times (see \cite{KBM2006}). This is utilized in \cite{RBM2019} to construct a consistent semi-parametric estimator of the input LST for a L\'evy-driven queue and an hypothesis testing procedure in \cite{MR2021}. This methodology is related to the generalized method of moments for continuous time Markov chains sampled at random times (see \cite{DG2004}). Remarkably, the estimators based on Poisson sampling are useful even if the observations times are deterministic and equidistant.  By resampling the observations the Poisson sampling process can be approximated, yielding a consistent estimator for the input parameters (see \cite{N2020}). This framework further has potential for developing non-parametric estimation techniques of the distribution $G$ by inversion of the LST estimator.

Establishing asymptotic normality of the estimation errors is typically possible, however, simulation techniques are often required to compute the asymptotic variance. Minimizing the asymptotic variance given periodic workload observations is an open question.  Some promising approaches include resampling schemes, MCMC and considering high-frequency sampling limits. Ultimately,  we would like to get as close as possible to the MLE efficiency, hence exploring the boundaries of existing and new estimation techniques is an important ongoing challenge.

\bibliographystyle{abbrv}
\bibliography{MG1_Inference}
\end{document}